\documentclass[conference]{IEEEtran}
\IEEEoverridecommandlockouts
\usepackage{cite}
\usepackage{amsmath,amssymb,amsfonts}
\usepackage{graphicx}
\usepackage{textcomp}
\usepackage{xcolor}

\usepackage{algorithm}
\usepackage[noend]{algpseudocode}
\usepackage{multirow} 

\def\BibTeX{{\rm B\kern-.05em{\sc i\kern-.025em b}\kern-.08em
    T\kern-.1667em\lower.7ex\hbox{E}\kern-.125emX}}

\begin{document}

\title{GPU Algorithm for Earliest Arrival Time Problem in Public Transport Networks}






\DeclareRobustCommand*{\IEEEauthorrefmark}[1]{
  \raisebox{0pt}[0pt][0pt]{\textsuperscript{\footnotesize\ensuremath{#1}}}}

\author{

\IEEEauthorblockN{
Chirayu Anant Haryan\IEEEauthorrefmark{1}, 
G. Ramakrishna\IEEEauthorrefmark{2}, Rupesh Nasre\IEEEauthorrefmark{3} and 
Allam Dinesh Reddy\IEEEauthorrefmark{4}}

\IEEEauthorblockA{\IEEEauthorrefmark{1}\IEEEauthorrefmark{2}\textit{Department of CSE}, \textit{Indian Institute of Technology Tirupati}, \textit{India} \\}

\IEEEauthorblockA{\IEEEauthorrefmark{3}\textit{Department of CSE}, \textit{Indian Institute of Technology Madras}, \textit{India}}

\IEEEauthorblockA{\IEEEauthorrefmark{4}\textit{Zippr Private Limited, Hyderabad, India}}

\IEEEauthorblockA{\IEEEauthorrefmark{1}cs18s503@iittp.ac.in, \IEEEauthorrefmark{2}rama@iittp.ac.in, \IEEEauthorrefmark{3}rupesh@iitm.ac.in, 
\IEEEauthorrefmark{4}dinesh.allam99@gmail.com}
}

\maketitle
\begin{abstract}

Given a temporal graph $G$, a source vertex $s$, and a departure time at source vertex $t_s$, the earliest arrival time problem (\textsc{eat}) is to start from $s$ on or after $t_s$ and reach all the vertices in $G$ as early as possible. Ni et al. have proposed a parallel algorithm for \textsc{eat}  and obtained a speedup up to $9.5$ times on real-world graphs with respect to the connection-scan serial algorithm by using multi-core processors. 

We propose a topology-driven parallel algorithm for \textsc{eat} on public transport networks and implement using general-purpose programming on the graphics processing unit (\textsc{gpu}). 
A temporal edge or connection in a temporal graph for a public transport network is associated with a departure time and a duration time, and many connections exist from $u$ to $v$ for an edge $(u,v)$.
We propose two pruning techniques \emph{connection-type} and \emph{clustering}, and use \emph{arithmetic progression} technique appropriately to process many connections of an edge, without scanning all of them.
In the connection-type technique, the connections of an edge with the same duration are grouped together.
In the clustering technique, we follow 24-hour format and the connections of an edge are partitioned into 24 clusters so that the departure time of connections in the $i^{th}$ cluster is at least $i$-hour and at most $i+1$-hour. The arithmetic progression technique helps to store a sequence of departure times of various connections in a compact way. We propose a hybrid approach to combine the three techniques connection-type, clustering and arithmetic progression in an appropriate way.
Our techniques achieve an average speedup up to $59.09\times$ when compared to the existing connection-scan serial algorithm running on \textsc{cpu}. Also, the average speedup of our algorithm is $12.48\times$ against the parallel edge-scan-dependency graph algorithm running on \textsc{gpu}.


\end{abstract}

\begin{IEEEkeywords}
Earliest arrival time, temporal graphs, public transport networks, parallel algorithms, graphics processing unit.
\end{IEEEkeywords}

\section{Introduction}
Route planning algorithms in road networks and public transport networks are well studied and are designed to find various kinds of paths such as shortest paths, fastest paths, and earliest arrival paths.
A public transport network primarily consists of a scheduled time table information (departure and arrival times) of public transport vehicles along various routes, and the underlying road network.
A public transport network can be modeled as a temporal graph that consists of vertices, edges and connections. The road network structure is captured by the topology of the graph, i.e., a road that starts from $u$ and ends with $v$ is represented using an edge from a vertex $u$ to a vertex $v$, and is denoted as $(u,v)$. Multiple vehicles going through a road $(u,v)$ with various departure times and arrival times are captured using connections; a connection $(u,v, t, \lambda)$ is a 4-tuple, which indicates that there is a road from $u$ to $v$, there is a vehicle whose departure time at $u$ is $t$ and the duration time of the vehicle from $u$ to $v$ is $\lambda$.
Since the information about edges is implicitly available in connections, a temporal graph is defined as a weighted directed graph consists of vertices and connections, where every connection is associated with a departure time and a duration time.
A path $P$ in a temporal graph is a \emph{time respecting path} if the departure time of the outgoing connection is at least the arrival time of the incoming connection at every intermediate vertex in $P$.
For a query consists of a source vertex $s$, destination vertex $d$ and a departure time $t_s$, a time respecting path from $s$ to $d$, whose departure time from $s$ is at least $t_s$ and the arrival time at $d$ is minimum, is referred to as \emph{earliest arrival path}, and the arrival time at $d$ is referred to as \emph{earliest arrival time}.
Given a temporal graph $G$, a source vertex $s$, and a departure time at source vertex $t_s$, the \textsc{earliest arrival time} (\textsc{eat}) problem of our focus in this paper is to find the earliest arrival times from a single source $s$ to the rest of the vertices \cite{IntriguinglyPaper}. 
In the goal-directed version of the earliest arrival time problem, earliest arrival time is computed from a source vertex to a designated destination vertex \cite{IntriguinglyPaper}. In case of profile-search problem, the set of non-dominated earliest arrival paths that depart in a specified time interval are to be computed \cite{IntriguinglyPaper}.
\textsc{eat} and their variants are well studied in not only on public transport networks, but also on real world temporal graphs such as facebook, flicker, dblp and arxiv using serial computing and multi-core computing~\cite{pathProblemsInTemporalGraphs,parallelAlgorithmForEAT}.

A variant of Dijkstra's algorithm is used to solve \textsc{eat} problem by storing a public transport network information using time-expanded model and time-dependent model \cite{foremostJourney2003}. Dijkstra's algorithm and their variants are vertex-centric algorithms, whereas the connection-scan algorithm (\textsc{csa}) proposed by Dibbelt et.al. used to solve \textsc{eat} is an edge-centric algorithm~\cite{IntriguinglyPaper}. This algorithm has become popular due to the simplicity, spatial data locality, and is versatile to solve multi-criteria profile queries and minimum expected arrival time problem~\cite{IntriguinglyPaper}. Later, Wu et.al. have used \textsc{csa} to  solve \textsc{eat} problem on real-world temporal graphs~\cite{pathProblemsInTemporalGraphs}. There after, \textsc{eat} on real world temporal graphs is solved using a parallel Edge-Scan-Dependency-Graph (\textsc{esdg}) algorithm on multi-core processors and obtain $9.5 \times$ speedup with respect to the connection-scan algorithm \cite{parallelAlgorithmForEAT}. 

Modern architectures such as multi-core central processing unit (\textsc{cpu}) and many-core graphics processing unit (\textsc{gpu}) are used heavily to exploit the various kinds of parallelism and the performance of various applications.  The number of cores in a multi-core \textsc{cpu}  typically varies from 2 to 128, whereas a many-core \textsc{gpu} consists of hundreds or thousands of cores. The architecture and design philosophy of multi-core \textsc{cpu} and many-core \textsc{gpu} are fundamentally different; \textsc{cpu}s are designed to decrease the latency time, whereas increasing the throughput is the main focus in \textsc{gpu}s \cite{gpuBook} and hence, the adaption of algorithms designed for multi-core architecture to \textsc{gpu} architecture is a non-trivial task. Various graph problems such as finding biconnected components and strongly connected components are explored using multi-core and many-core architectures \cite{biconnectedComponents},\cite{stronglyConnectedCompoenents}.  
To the best of our knowledge, there is no work on \textsc{eat} problem using  general purpose graphics processing units \textsc{(gpgpu)}.  We propose parallel algorithms to solve \textsc{eat} in public transport networks, on many-cores using \textsc{gpgpu}.

We now state our main contributions.
\begin{itemize}
 \item \emph{Data-structures:} We exploit the temporal information in public transport networks such as departure time and duration time on every connection and propose various data-structures to retrieve useful connections efficiently.
 \item \emph{Algorithms:} We get inspiration from connection-scan algorithm and propose a few topological driven parallel algorithms to retrieve useful connections and ignore the rest of them for solving \textsc{eat}.
 Vertex-centric and edge-centric are well known approaches in algorithmic graph theory. A variant of edge-centric, called as \emph{connection-type centric} is proposed in our algorithms, in which many connections of same type are processed together. 
 \item \emph{Data-enhancement:} We exploit the spatial information of trips (a sequence of locations that are traversed by a vehicle in a public transport system), and propose to add new artificial connections.
 Our data-enhancement technique called \emph{sub-trips} approximately doubles the speedup of all the proposed algorithms. 


 \item \emph{Speedup:} We have implemented the proposed algorithms on \textsc{gpu} and run various experiments on popular public-transport data sets such as London, Los Angeles, Paris, Sweden and Switzerland. 
 With respect to the connection scan algorithm on \textsc{cpu}, the average speedup is between $2.29\times$ and $59.09\times$, whereas the average speedup is between $1.63\times$ and $12.48\times$ when compared to the parallel \textsc{esdg} algorithm on \textsc{gpu}.
 
\end{itemize}

\subsection{Preliminaries}

We use $G=(V,C)$ to denote a temporal graph, where $V$ and $C$ are vertices and connections in $G$, respectively. A connection is a quadruple of ($u$, $v$, $t$, $\lambda$), where there exists a public transport vehicle that starts from $u$ at departure time $t$ and reaches $v$ at $t+\lambda$.  
For a temporal graph $G$, the \emph{temporal diameter} of $G$, denoted by $d(G)$, is defined as the maximum number of connections in an earliest arrival path from $s$ starting with departure time $t_s$ to any reachable vertex from $s$, over all vertices $s$ in $G$ and over all times $t_s \geq 0$.
In public transport data, the departure time of a temporal edge is represented in seconds. However, we use \texttt{HH:mm} or \texttt{HH:mm:ss} format a few times to illustrate certain concepts, where \texttt{HH}, \texttt{mm} and \texttt{ss} denotes the hours, minutes, and seconds of a time stamp, respectively. For example, \texttt{13:20:34} denotes 13 hours, 20 minutes and 34 seconds in a 24-hour format.
The state-of-art serial and parallel algorithms to solve single source earliest arrival path problem on temporal graphs are connection-scan algorithm and edge-scan-dependency-graph (\textsc{esdg}) algorithm, respectively, and are described below. 

\noindent
\textbf{Connection-Scan Algorithm:} 
Given a graph $G(V,C)$ and a query $(s,t_s)$ the algorithm calculates the earliest arrival times in $O(|C|)$ time. The algorithm begins with an initialization phase in which the earliest arrival time of all vertices except the source is set to $\infty$ and earliest arrival time of $s$ is set as $t_s$. Further connections in $C$ are relaxed in non-decreasing order of departure time, as shown in Algorithm~\ref{algorithmConnectionScanSerial}. The final earliest arrival times are acquired when all the connections are relaxed.

\begin{algorithm}[H]

\textbf{Input:} $G=(V,C)$ and Query $(s,t_s)$. \\
{\small/*\texttt{ $C$ is a sequence of connections in $G$ that are arranged  in non-decreasing order based on their departure time.} */}\\
\textbf{Output:} Earliest Arrival Times for all vertices in $G$. 
\caption{Connection-Scan Algorithm\label{algorithmCSA}}
\label{algorithmConnectionScanSerial}
\begin{algorithmic}[1]
\ForAll{vertex ${u \in  V\setminus s}$}
\State{${e[u]}$ = $\infty$}
\EndFor
\State{$e[s]$ = $t_s$}
\For{\textbf{each }($u$, $v$, $t$, $\lambda$) $in$ $C$} 
\If{($e[u] \leq t$  and $t$ + $\lambda$ $<$ $e[v]$)} \label{lineRelax1}
\State $e[v]$ = $t$ + $\lambda$ \label{lineRelax2}
\EndIf
\EndFor
\end{algorithmic}
\end{algorithm}

\noindent
\textbf{Edge-Scan-Dependency-Graph Algorithm:}
Ni et al. have designed a parallel algorithm to solve \textsc{eat} by introducing \emph{edge-scan-dependency graph} (\textsc{esdg})  \cite{parallelAlgorithmForEAT}. 
An edge-scan-dependency graph $\tilde{G}$ of a temporal graph $G$ is defined as follows.
The connections in $G$ are treated as vertices in $\tilde{G}$.
Let $c=(u,v,t,\lambda)$ and $c'=(u',v',t',\lambda')$ be two connections in $G$. Then, there exists a directed edge from $c$ to $c'$ if they satisfy the following two conditions:

    1. $v = u'$ and $t' \geq t+\lambda$

    2. $\nexists$ $(u'',v'',t'',\lambda'')$ $\in$ $C$ such that $t'' \geq t+\lambda$ and $t''+\lambda'' < t'+\lambda'$.

Note that the edge-scan-dependency graph $\tilde{G}$ is a directed acyclic graph.
For each vertex $v$ in $\tilde{G}$, $level(v)$ is defined as follows: if there are no incoming edges to $v$, then \mbox{$level(v)=0$};  otherwise, $level(v)$ is equal to the length of a longest path from a vertex whose level is zero to $v$.
In the parallel edge-scan-dependency graph algorithm, 
all the connections within the same level are relaxed 
(Lines~\ref{lineRelax1} and \ref{lineRelax2} in Algorithm~\ref{algorithmCSA})
in parallel, and the connections in different levels are
relaxed in increasing order of their level number. 

\noindent
\textbf{Arithmetic Progression Technique:} We use the following technique in our algorithm, to represent a sequence $S$ of positive integer in  compact way \cite{frequencyBasedPaper}. In each iteration, an uncovered smallest number $a \in S$ is chosen, and finds a longest arithmetic progression starting with $a$ and cover maximum number of uncovered numbers. The same step is repeated until all the numbers in $S$ are covered. Finally, each arithmetic progression can be compactly represented using the first term, last term and the difference between the consecutive terms in the arithmetic progression. For instance, the arithmetic progression (10,15,20,25,30,35) can be compactly represented as (10,35,5). 


\section{Proposed Incremental Parallel Algorithms}

In this section, we describe our parallel algorithms that are developed with incremental improvements.
We first propose a topological driven connection-version algorithm that is inspired from connection-scan algorithm.
Further, we describe pruning techniques connection-type, arithmetic progression, and clustering. Later, we combine these techniques to obtain a hybrid technique Cluster-\textsc{ap}. Finally, we illustrate a warp-based algorithm and data-enhancement technique.

For each vertex $u \in V$, we use $e[u]$ in our algorithms to store the earliest arrival time from the source vertex to $u$.
Our proposed algorithms use two subroutines, namely \textsc{Initialize} and \textsc{Relax}.
In the initialization subroutine, shown in Algorithm \ref{algorithmInitialize}, $G$ is a temporal graph, $s$ is a source vertex and the $t_s$ is the departure time at $s$. 
This subroutine sets the earliest arrival times for all vertices except the source vertex as $\infty$ in parallel. The earliest arrival time of source vertex is set as $t_s$. Also, we maintain an array $active$[] of $|V|$ Boolean flags and set true or false in each location to denote that the corresponding vertex is active or passive, respectively.  The initialization routine sets all the vertices except the source as passive in parallel. The source vertex status is set as active.



\begin{algorithm}[H]
\caption{\textsc{Initialize}$(G,e,s,t_s)$}
\label{algorithmInitialize}
\begin{algorithmic}[1]
\ForAll{vertex ${u \in  V}$} \textbf{in parallel}
\State{${e[u]}$ = $\infty$}
\State{$active[u]$ = $false$}
\EndFor
 \State{$e[s]$ = $t_s$}
 \State{$active[s]$ = $true$}
\end{algorithmic}
\end{algorithm}
\vspace*{-10pt}
\begin{algorithm}[H]
\caption{\textsc{Relax}($u$, $v$, $t$, $\lambda$)}
\label{algorithmRelax}
\begin{algorithmic}[1]
\If{($e[u]$ $\leq$ $t$  and $t$ + $\lambda$ $<$ $e[v]$)}
\State $e[v]$ = $t$ + $\lambda$
\State{$active[v]$ = $true$}
\State{\textbf{return} $true$}
\Else
\State{\textbf{return} $false$}
\EndIf
\end{algorithmic}
\end{algorithm}

The \textsc{Relax} subroutine described in Algorithm \ref{algorithmRelax} receives a connection $(u,v,t,\lambda$), and finds whether the connection $(u, v, t, \lambda)$ reduces the earliest arrival time of $v$.
The \textsc{Relax} subroutine identifies whether the connection is useful by comparing $t$ against $e[u]$, and updates $e[v]$ with $t + \lambda$ if $e[u]$ $\leq$ $t$ and $t+\lambda < e[v]$. Further, $active[v]$ is set as true and true is returned in case $e[v]$ is decreased, otherwise false is returned.


\subsection{Connection-Version}

The key idea in this parallel algorithm is to provide a one-to-one mapping between threads and connections, and each thread relaxes the connection it is mapped to in parallel. At the beginning of the algorithm, the initialization subroutine shown in Algorithm~\ref{algorithmInitialize} is called.
Further, in each iteration, $i^{th}$ thread considers $i^{th}$ connection $(u,v,t,\lambda)$. If $u$ is active, $e[u] \leq t$, and $t+\lambda < e[v]$, then $e[v]$ is replaced with $t + \lambda$,  $v$ is marked as active, as per the definition provided in \textsc{Relax} subroutine. Further, $u$ is made as passive. All the connections whose starting vertex is $s$ are processed in the first iteration, since $s$ is the only active vertex in the first iteration. In all the subsequent iterations, the connections originated from active vertices are processed and propagates earliest arrival times to the other end vertices. We perform multiple iterations of relaxing connections until there is no change in the earliest arrival times of any vertex from the previous iteration. The formal description of algorithm for connection-version is given in Algorithm \ref{algoConnectionVersion}. In Algorithm~\ref{algoConnectionVersion}, threads relax connections in parallel, and hence the earliest arrival times at the end of the first phase need not be correct. However, the correctness of the algorithm is maintained as we perform multiple iterations. 
For an earliest arrival path in $G$, from $s$ to an arbitrary vertex $x$ in $V(G)$, having $k$ connections,  Algorithm~\ref{algoConnectionVersion} maintains a loop invariant that, at the end of $k^{th}$ iteration, $e[x]$ is equal to the earliest arrival time of $x$. The numbers of iterations in this algorithm in the worst case is $d(G)$, and thus the running time of the algorithm is $O(d(G))$.

\begin{algorithm}[H]
\textbf{Input:} $G=(V,C)$ and Query$(s,t_s)$.\\
\textbf{Output:} Earliest Arrival Times of all vertices in $G$. 
\caption{Connection Version}
\label{algoConnectionVersion}

\begin{algorithmic}[1]
\State{\textsc{Initialize}$(G,e,s,t_s)$}\\
flag = $true$
\While{flag}
\State{flag = $false$}
\ForAll{$(u,v,t,\lambda)$ $\in$ $C$}\textbf{ in parallel}
 \If{$active[u]$}
 \State flag = flag \textbf{or} \textsc{Relax}$(u,v,t_c,\lambda)$ 
 \State{$active[u]$ = $false$}
\EndIf
\EndFor
\EndWhile
\end{algorithmic}
\end{algorithm}


\subsection{Connection-Type-Version}
We observe that only a few connections in each iteration influence the earliest arrival times in the previous algorithm. 
Based on this observation, we launch threads for handling only useful connections and ignore the rest of them, in this version.
An important idea in this algorithm is to prune several connections that are not relevant. We introduce a data structure \emph{connection type} to achieve pruning. During the pre-processing time, we  partition the connections in $C$ into connection types, based on the following equivalence relation $R$ on $C$:  ${(u_a,v_a,t_a,\lambda_a)R(u_b,v_b,t_b,\lambda_b)}$ if and only if ${(u_a,v_a,t_a,\lambda_a),(u_b,v_b,t_b,\lambda_b) \in C}$,  ${u_a = u_b}$, ${v_a = v_b}$ and ${\lambda_a = \lambda_b.}$ In simple words, the connections whose endpoints and duration are the same, belong to the same connection type.  We use the notation $C_{u,v,\lambda}$ to represent a connection type, i.e.,  $C_{u,v,\lambda}$ denotes the set of connections from $u$ to $v$ whose duration is $\lambda$. In each connection type, all the connections are sorted according to their departure time in pre-processing time.
We use \textsc{getConnection} function to obtain a first connection ${(u,v,t_c,\lambda)}$ from $C_{u,v,\lambda}$, whose departure time is at least the earliest arrival time of $u$, i.e.,  $t_c = \min\big\{ t \mid (u,v,t,\lambda)\in C_{u,v,\lambda}\wedge t \geq e[u]\big\}$, using linear search.

We have one-to-one mapping between connection types and threads in this algorithm, i.e., each thread is responsible to process the connections in a connection-type. At the beginning, the steps given in the initialization routine are executed. In each iteration, $i^{th}$ thread associated with $i^{th}$ connection-type $C_{u,v,\lambda}$, checks whether the vertex $u$ is active. If $u$ is active, the thread identifies a connection  ${(u,v,t_c,\lambda)}$ using \textsc{getConnection}$(C_{u,v,\lambda})$ and the identified connection is relaxed using \textsc{Relax()}. Multiple iterations are executed until there is no change in the earliest arrival times in consecutive iterations.
Relaxation of any connection ${(u,v,t,\lambda)}$ $\in$ $C_{u,v,\lambda}$ such that ${t > t_c}$, does not improve $e[v]$, because all the connections in a connection type have same duration. Thus pruning such connections in \textsc{getConnection} improves the running time without disturbing the correctness of the algorithm.


\begin{algorithm}[H]
\textbf{Input:} Connection Types of $G$ and Query($s,t_s$).\\
\textbf{Output:} Earliest Arrival Times of all vertices in $G$. 
\caption{Connection-Type-Version}
\label{algoConnection-type}
\begin{algorithmic}[1]
\State{\textsc{Initialize}$(G,s,t_s)$}\\
flag = $true$
\While{flag is $true$}
\State{flag = $false$}
\ForAll{Connection Type ${C_{u,v,\lambda}}$} \textbf{in parallel}
\If{$active[u]$}
\State{$(u,v,t_c,\lambda)$ = \textsc{getConnection}$(C_{u,v,\lambda})$} \label{lineGetConnection}
\State flag = flag \textbf{or} \textsc{Relax}$(u,v,t_c,\lambda)$ 
\State{$active[u]$ = $false$}
\EndIf
\EndFor

\EndWhile
\end{algorithmic}
\end{algorithm}

    \subsection{Connection-Type-\textsc{ap}-Version}
We obtain an improvement to the previous algorithm by combining the idea of connection-type with arithmetic progression.
The main ingredient in this algorithm is to represent all the connections in every connection type in a compact form using arithmetic progression (\textsc{ap}).
This approach is based on the observation that departure times of buses and trains often follow a common pattern. 
 For a moment, let us assume that the departure times of  connections belong to the same connection-type are \texttt{8:00}, \texttt{8:15}, \texttt{8:30, \ldots,18:00};  As this sequence follows an arithmetic progression, all the departure times can be represented using a single \textsl{\textsc{ap}} \emph{tuple} with three terms as $(8:00,18:00,15)$, indicating there are connections having departure times from \texttt{8:00} to \texttt{18:00} for every $15$ minutes. In general, the departure times of all the connections in a connection type need not follow an \textsc{ap}. However,  all the connections in a connection type  $C_{u,v,\lambda}$ can be represented compactly by using a set  $T_{u,v,\lambda}$ of \textsc{ap} tuples \cite{frequencyBasedPaper}. For every connection $(u,v,t,\lambda) \in C_{u,v, \lambda}$, there exist at least one \textsc{ap} tuple $(startTime, endTime, difference) \in T_{u,v,\lambda}$, such that  ${t = startTime + i \times difference}$, where $i \in \{0,1,2,\ldots,k\}$ and ${k = \tfrac{endTime - startTime}{difference}}$.  Also, note that the expansion of all \textsc{ap} tuples in $T_{u,v,\lambda}$ results in  $C_{u,v,\lambda}$ without any additional departure times. All the \textsc{ap} tuples in every connection type are arranged in increasing order with respect to their first term, in preprocessing time. Given a set $T_{u,v,\lambda}$ of \textsc{ap}-tuples that corresponds to a connection-type, the \textsc{getConnectionFromAPs} algorithm, shown in Algorithm~\ref{getConnectionFromAP}  identifies an  useful connection based on the earliest arrival time of $u$.  In \textsc{getConnectionFromAPs} algorithm,  from each \textsc{ap}, we consider the  first departure time which is at least $e[u]$, and finds a minimum departure time among them.




Now, we shall look at Connection-type-\textsc{ap} algorithm. The key task of each thread in this algorithm is to  process all the \textsc{ap} tuples in a connection-type and identifies a useful connection, using Algorithm~\ref{getConnectionFromAP}. 
The steps in Connection-type-\textsc{ap} algorithm are identical to the steps in Algorithm~\ref{algoConnection-type} except Line~\ref{lineGetConnection}.  In this approach, \textsc{getConnection} function in  Line~\ref{lineGetConnection} of Algorithm~\ref{algoConnection-type} is replaced with \textsc{getConnectionFromAPs} function, which is described in Algorithm~\ref{getConnectionFromAP} .



	
\begin{algorithm}
\textbf{Input:} ${u,v,\lambda,T_{u,v,\lambda}}$ \\
{\small/*\texttt{ $u$ and $v$ are two vertices in $G$ and $\lambda$ is the duration of the connections in $C_{u,v,\lambda}$.} */} \\
{\small/*\texttt{ $T_{u,v,\lambda}$ is set of \textsc{ap} tuples in connection type $C_{u,v,\lambda}$.} */}\\
\textbf{Output:} ${(u,v,t_c,\lambda)}$ 
\caption{\textsc{getConnectionFromAPs}}
\label{getConnectionFromAP}
\begin{algorithmic}[1]
\State{$t_c$ = $\infty$}
\ForAll{($startTime$, $endTime$, $difference$) $\in$ ${T_{u,v,\lambda}}$}
\If{$startTime < e[u] \leq {endTime}$} 
\State{${i = \lceil{\frac{e[u] - startTime}{difference}}\rceil }$ }
\State{${t_c}$ = ${min(t_c,startTime + i \times difference)}$}
\EndIf
\If{$e[u]\leq startTime$}
\State{${t_c=min(t_c,startTime)}$}
\EndIf
\EndFor
\State{\textbf{return} ${(u,v,t_c,\lambda)}$ }
\end{algorithmic}
\end{algorithm}

    \subsection{Cluster-\textsc{ap}-Version}
We propose a hybrid technique to search a connection efficiently, using three ideas connection-type, arithmetic progressions and clustering.
The main idea in this algorithm is to partition all the connections belong to every connection type into 24 parts, where each part represents an hour of the day. Further, we represent all the connections in each part in a compact format using an arithmetic progression, which is described in the previous section. We  partition all the connections in each connection-type  $C_{u,v,\lambda}$ into clusters $C[0]_{u,v,\lambda}, \ldots, C[23]_{u,v,\lambda}$, where $C[i]_{u,v,\lambda}$ denotes the set of connections whose departure time  $t \in$\texttt{[$i$:00:00, $i$:59:59]}.
Let $T[i]_{u,v,\lambda}$ denote the set of \textsc{ap}-tuples correspond to the connections in $C[i]_{u,v,\lambda}$ . Partitioning all the connections into connection-types based on their duration time, and then further partitioning them into 24 clusters based on the departure time, and finally representing them using arithmetic progression happens during the prepossessing time. 

Now, we shall look at the main steps in this algorithm. The number of threads that we launch in this algorithm is equal to the total number of connection types. Each thread that indents to process the connections in $C_{u,v,\lambda}$, extracts the hour information $k$ from $e[u]$, i.e., $k=e[u]/3600$ and then process all the \textsc{ap}-tuples in $T[k]_{u,v,\lambda}$.
Suppose the $k^{th}$ part does not have any connections, then the first connection from the next non-empty part is treated as an useful connection. The correctness of the algorithm is not affected with this heuristic, as the connections whose departure time is at least $e[u]$ are not ignored. In practice, the running time is improved as the connections belong to all the previous hours of $e[u]$ are ignored.

     \subsection{Edge-Version}
     In all the approaches illustrated so far, every thread relaxes at most one connection in an iteration. In this approach, we explore the behavior by decreasing the number of threads to be launched and increasing the amount of work per thread, i.e., multiple connections are  relaxed by a thread. 
     In the pre-processing time, all the connections are partitioned into edges, i.e., all the connections starting from a vertex $u$ and ending with a vertex $v$ are associated with an edge $(u,v)$. Further the connections associated within each edge are grouped into different connection-types based on their duration time and clustered them based on their departure time.
     
     We have a one-to-one mapping between threads and edges in the edge-version algorithm. 
     At the beginning, we perform initialization as given in Algorithm \ref{algorithmInitialize}. Then each thread, process all  the connections associated with an edge efficiently. In particular,  each thread explores all the connection types on an edge, identify a single useful connection from each connection-type using Cluster-\textsc{ap} approach and relaxes the identified connections. All the edges are processed for multiple iterations  until there is no change in the earliest arrival times in consecutive iterations.

     
     

     
    \subsection{Warps-Version}
\label{warps}
   When a kernel call is made, multiple blocks of threads are assigned to a streaming multiprocessor (\textsc{sm}) in a \textsc{gpu}. 
   These  blocks  are  further  divided  into \emph{warps}, where each warp is a group  of  32  threads, and all  the  threads in each warp  execute  the  same instruction  at  any  point  in  time.
   Each thread has its own instruction counter and hence it is possible for a thread to choose different execution path than the other threads, this phenomenon is called thread divergence. Thread divergence within a warp, severely affects the execution time. Also, the thread divergence is more likely to happen in graph algorithms, and so in the proposed algorithms. The divergence of a thread in our algorithms depends on the source vertex of the connection/connection-type/edge associated with the thread. A thread terminates its processing if the corresponding source vertex is passive, otherwise, the processing is continued. Also, a memory access is required to know whether a vertex is active or passive.
   The end vertices of connections being processed by the threads of a warp need not be same, and hence the memory accesses are scattered. These scattered memory accesses add a penalty to the execution time. We propose a mechanism to reduce the thread divergence and scattered memory accesses.
   

    In this algorithm, we keep a one-to-one mapping between warps and edges. The algorithm begins with executing the steps given in the initialization routine.  We use all the threads in a warp to process all the connection-types of an edge.
We divide the connection-types $ct_0, ct_1, \ldots, ct_k$ in an edge into $\lceil k/32 \rceil$ groups, so that group-$i$ consists of connection-types $ct_{32\times i}, \ldots, ct_{32\times i + 31}$, where $i \in \{0, \lceil k/32 \rceil-1\}$, and the last group has at most 32 connection-types. Further, we iterate all the threads in a warp for $\lceil k/32 \rceil$ times, where the connection-types in the $i^{th}$ group are processed in $i^{th}$ iteration.
The organization of connections in connection-types is the same as that of in the Cluster-\textsc{ap} approach, and hence the process of relaxing the connections in a connection-type is same. Multiple iterations are performed until the earliest arrival times do not change in consecutive iterations.
    
    We recall that all the connection-types of an edge share a same source vertex. Also, all the connection-types of an edge are processed by threads belonging to the same warp.
    All the threads in a warp terminate if their common source vertex is inactive; otherwise, all of them proceed to process the connections in a connection-type. Thus, the thread divergence is reduced and the scattered memory accesses are eliminated.
   
\subsection{Sub-trips} \label{subtrips}
    
        
        The focus of all parallel algorithms described so far is to reduce the run-time complexity of each iteration, where the number of iterations in all these algorithms is equal to the temporal diameter. In this approach, we reduce the temporal diameter, subsequently the number of iterations of the algorithm, by adding few artificial edges in an organized way, during the preprocessing time. In a public transport system, each vehicle goes through a sequence of bus-depots, and typically the scheduled departure times and arrival times at each bus-depots are known.  We exploit this information to reduce the temporal diameter.

    A \emph{trip} in a temporal graph is a sequence of connections connecting a sequence $(v_1,\ldots,v_k)$ of vertices,  such that for every two consecutive connections $(v_i,v_{i+1},t_i,\lambda_i)$ and $(v_{i+1},v_{i+2},t_{i+1},\lambda_{i+1})$,  $t_i + \lambda_i \leq t_{i+1}$, where  $1 \leq i \leq k-2$.
    A sub-path of a trip is referred to as \emph{sub-trip}.
    We divide each trip into sub-trips of non-overlapping sequence of connections, and add artificial connections between the end vertices for each sub-trip as follows:     for  a sub-trip on a sequence  $(v_i,v_{i+1},t_i,\lambda_i)$, $(v_{i+1},v_{i+2},t_{i+1},\lambda_{i+1}), \ldots,$ $(v_j,v_{j+1},t_j,\lambda_j$) of connections, where $i<j$,  we add an artificial connection between $v_i$ and $v_{j+1}$, whose departure time is $t_i$ and duration time is $t_j + \lambda_j - t_i$.  These new connections work as short-cuts and help to reduce the temporal diameter  of the graph.
   
    We reduce the temporal diameter of a temporal graph, by dividing each trip of length $k$ into $\frac{k}{r}-1$ sub-trips of length $r$,  and a last sub-trip of length $k\%r$. We choose $r = \sqrt{k}$, to minimize the number of sub-trips in each trip along with the number  of connections in each sub-trip.  All the parallel algorithms described in this paper can be run on the updated graph. Note that the sequence of connections in a trip are traversed by the same vehicle and the sequence of connections in all the trips are available in the input data-sets in General Transit Feed Specification \textsc{gtfs} format. The connection sequences of all these trips are being used to create sub-trips.

\section{Implementation Details and Optimizations}
In this section, we first describe a hierarchical representation to store temporal graphs for handling public transport networks and provide a preprocessing algorithm to obtain the hierarchical representation. We further propose implementation details to avoid read/write conflicts and handle race conditions. Finally, we conclude this section with a few heuristics that support to terminate our algorithm at an early stage.

\subsection{Graph Representation - Preprocessing Algorithm}

\begin{figure}
    \centering
    \includegraphics[width = 0.45\textwidth, keepaspectratio]{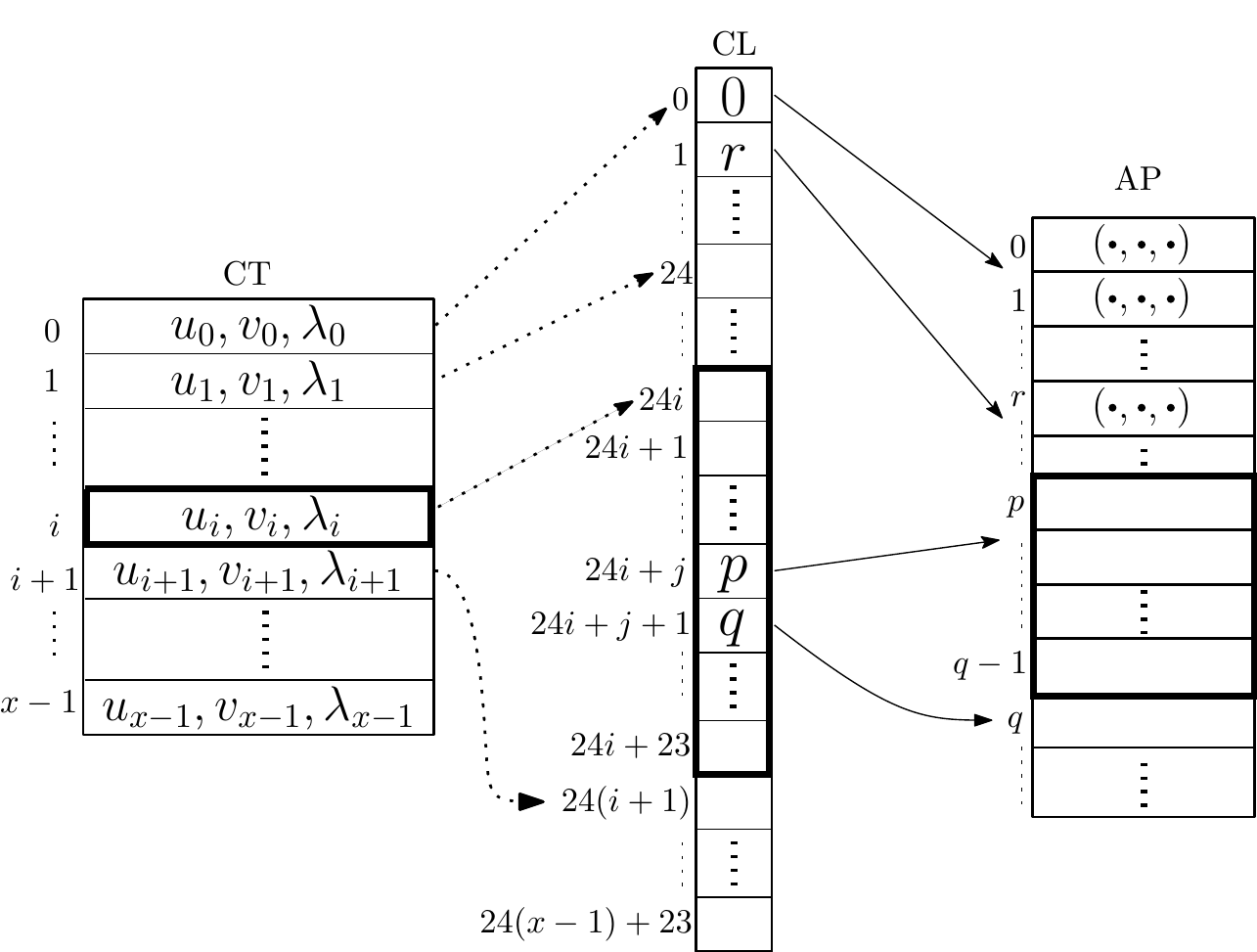}
    \caption{\label{fig:Cluster AP Data Representation} Data Representation}
    \vspace*{-10pt}
\end{figure}


We inspire from the compressed sparse row (CSR) representation and propose the following representation to store the connections of a temporal graph $G$ in a hierarchical manner. Three arrays $CT$[], $CL$[] and $AP$[] are used in this representation for efficient searching and are shown in Fig.~\ref{fig:Cluster AP Data Representation}.
Let $x$ denote the number of connection-types in $G$.
and $y$ denote the maximum number of clusters per connection-type over all the connection-types in $G$. 
For simplicity, $y$ is assumed as 24 in this illustration and this is configured in our implementation to handle various data-sets having more than 24 clusters.
 We use the leftmost array $CT$[], so that for each $i \in \{0,x-1\}$, the end vertices and the duration of $i^{th}$ connection-type in $G$ are stored in $CT[i]$. The rightmost array $AP$[] is used to store  the arithmetic progressions associated with all the clusters of all  connection-types. The intermediate array $CL$[] holds the starting index of the first AP from the $j$th cluster of $i$th connection-type;
i.e., for each $ i \in \{0, \ldots,x\}$, and for each $j \in \{0,\ldots,23\}$, the APs from $j^{th}$ cluster of $i^{th}$ connection-type are available in the range $[CL[24*i+j], CL[24*i+j+1])$ of positions in $AP$. 

Now, we shall look the preprocessing algorithm to obtain the hierarchical representation from $G$. For each edge $e$ in $G$, all the connections of $e$ are grouped into connection-types if their end vertices and duration are same; further all the connections in each connection-type are partitioned into clusters, so that the connections whose departure time $t \in$ \texttt{[j:00:00, j:59:59]} are placed in cluster-$j$; further arithmetic progressions are obtained to cover all the connections in each cluster of each connection-type \cite{frequencyBasedPaper}.
For each $i \in \{0,x-1\}$, and for each $j \in \{0, y-1\}$, we store the $r$ APs of cluster-$j$ of connection-type-$i$ consecutively in $AP$[] in increasing order, based on their first-term and the position of the first AP among $r$ APs is stored in $CL[24*i+j]$.

\subsection{Avoiding read/write conflicts}
In the connection version algorithm, a thread $t_i$ relaxes a connection $c_i = (u,v,t \lambda)$ if $active[u]$ is true and update $active[u]$ with false. The number of threads launched in this version is equal to the number of connections. However, all the threads need not execute the kernel code (Lines 6 to 8 in Algorithm~\ref{algoConnectionVersion}),  synchronously. 
Suppose there are two threads $t_1$ and $t_2$, associated with two different connections emerging from the same active vertex $u$, and $t_2$ is scheduled to run after $t_1$ completes.
In such a scenario, $t_2$ does not execute Lines 7 and 8,  because $active[u]$ is set to false by $t_1$, which is an incorrect behaviour. 
This scenario happens because $t_1$ writes in $active[u]$ before before $t_2$ reads from $active[u]$.
Hence, we use two arrays, namely $active$ and $nextactive$ to maintain the correctness of the algorithm.  The entries in the $nextactive$ are made $false$  in the beginning of each iteration, 
and then all the threads are launched. Then, a thread $t_i$ relaxes a connection $(u,v,t , \lambda)$ and updates $nextactive[v]$ with true if the earliest arrival time of $v$ is updated. 
After the iteration is completed, the values in the $nextactive$ array are copied to the $active$ array. Thus this technique avoids read/write conflicts and is used in the implementations of all the proposed parallel algorithms.


\subsection{Atomic operation}

In the proposed parallel algorithms, multiple threads may try to update the earliest arrival time for the same vertex. For instance, two threads associated with the connection $(u,v,t,\lambda)$ and $(x,v,t,\lambda)$ want to update the $e[v]$ to $3$ and $6$ respectively. Because of the race condition, the updated value of $e[v]$ can be either of one. If $e[v]$ is updated as $3$, the correctness is maintained, whereas the correctness is not maintained in the other case. 
This is possible, because 
$active[u]$ is made as false, and $active[u]$ need not become true again. Thereby,
the thread corresponding to connection $(u,v,t,\lambda)$ need not update $e[v]$ with $3$. Hence, we use atomicMin operation supported by \textsc{cuda}, to update the earliest arrival times and thus eliminate the interleaving execution by multiple threads. The atomic operation $atomicMin(address,val)$  updates the content in the $address$ with $val$, if $val$ is less than the content in the $address$. 
If we do not make $active[u]$ as false, then the connections are relaxed for multiple times. Hence earliest arrival times can be computed correctly without using atomic operation. However, we notice that the number of kernel launches and the running time increase in practice.

\subsection{Early Terminations of threads}
It can be recalled that the connections in a connection-type are arranged in increasing order based on their departure times. In Algorithm~\ref{algoConnection-type}, when a thread processes a connection-type $C_{u,v,\lambda}$, if the departure time of the first connection is greater than $e[v]$, then we terminate the thread as it cannot yield better value for $e[v]$. Also, if $e[u]$ is greater than the departure time of the last connection in $C_{u,v,\lambda}$, the thread terminates as there are no useful connections.


In the connection-type \textsc{ap} version, the departure times of the connections in a connection-type are represented as a sequence of \textsc{ap} tuples, such that  the first term of the \textsc{ap}s are in sorted order. When a thread processes a connection type $C_{u,v,\lambda}$, if the earliest arrival time of $v$ is lesser than the first term of the first \textsc{ap}, then the thread terminates as it cannot yield better earliest arrival time of $v$. 
We follow similar techniques in the Cluster-\textsc{ap}, edge-version and warps as the data representation is similar to connection-type \textsc{ap}.


\section{Experiments}

In this section, we evaluate our \textsc{gpu} based algorithms using the speedup achieved over the \textsc{cpu} based connection-scan algorithm, and the \textsc{gpu} based \textsc{esdg} algorithm. Also, we  provide few insights to increase the speedup, by fine tuning few parameters in the proposed algorithms. 

The machine being used in our experiments comprises of 1.62 \textsc{gh}z \textsc{nvidia geforce gtx} 1080 \textsc{t}i \textsc{gpu} with 11 \textsc{gb} of global memory and 128 cores in each of the 28 \textsc{sm}s,  an \textsc{intel xeon} E5-2620 v4 \textsc{cpu} running at 2.10\textsc{gh}z with 31 \textsc{gb} of main memory.
\textsc{cuda} version 9.0.176 and \textit{gcc} version 5.4.0 are used to implement parallel and serial algorithms, respectively.

We have used nine different public transport network data-sets for our experiments. The statistics for each of the data-set is given in  Table~\ref{data-set statistics}. The time table information is considered for more than 24 hours for certain data-sets based on the availability.
The last column in Table~\ref{data-set statistics} denotes the number of clusters in a data-set of size one hour.
\begin{table}
\caption{Data-set Statistics ($K = 10^3$, $M=10^6$) \label{data-set statistics}}
\begin{center}
\scalebox{0.9}
{
    \begin{tabular}{|l|c|c|c|c|c|}
    \hline
    \textbf{Data Set} & \textbf{Vertices} & \textbf{Edges} & \textbf{Connections} & \textbf{\begin{tabular}[c]{@{}c@{}}Connection\\ Type\end{tabular}} & \textbf{\begin{tabular}[c]{@{}c@{}}Clusters\\ 1Hr\end{tabular}} \\ \hline
    London & 20.8K& 25.5K & 14.06M &140.7K &26\\ \hline
    Paris &411 &1.56K &1.06M &3.08K &45\\ \hline
    Petersburg &7.6K &10.6K &4.43M &14.8K &49\\ \hline
    Switzerland &29.9K &74.1K &9.26M &102.6K &48\\ \hline
    Sweden &45.7K &101.9K &6.56M &158.3K &37\\ \hline
    New-York & 987 & 1.1K & 514.39K&1.9K&28 \\ \hline
    Madrid & 4.7K & 6.1K &1.99M&167.7K& 32  \\ \hline
    Los Angeles &13.9K &16K &1.97M &31.59K &30\\ \hline
    Chicago & 240 & 822 &98.15K&1.7K&27\\ \hline
\end{tabular}
}
 \vspace*{-10pt}
\end{center}
\end{table}

\subsection{Speedup - Parallel Factor }
\begin{figure}
    \centering
    \includegraphics[width=0.46\textwidth,keepaspectratio]{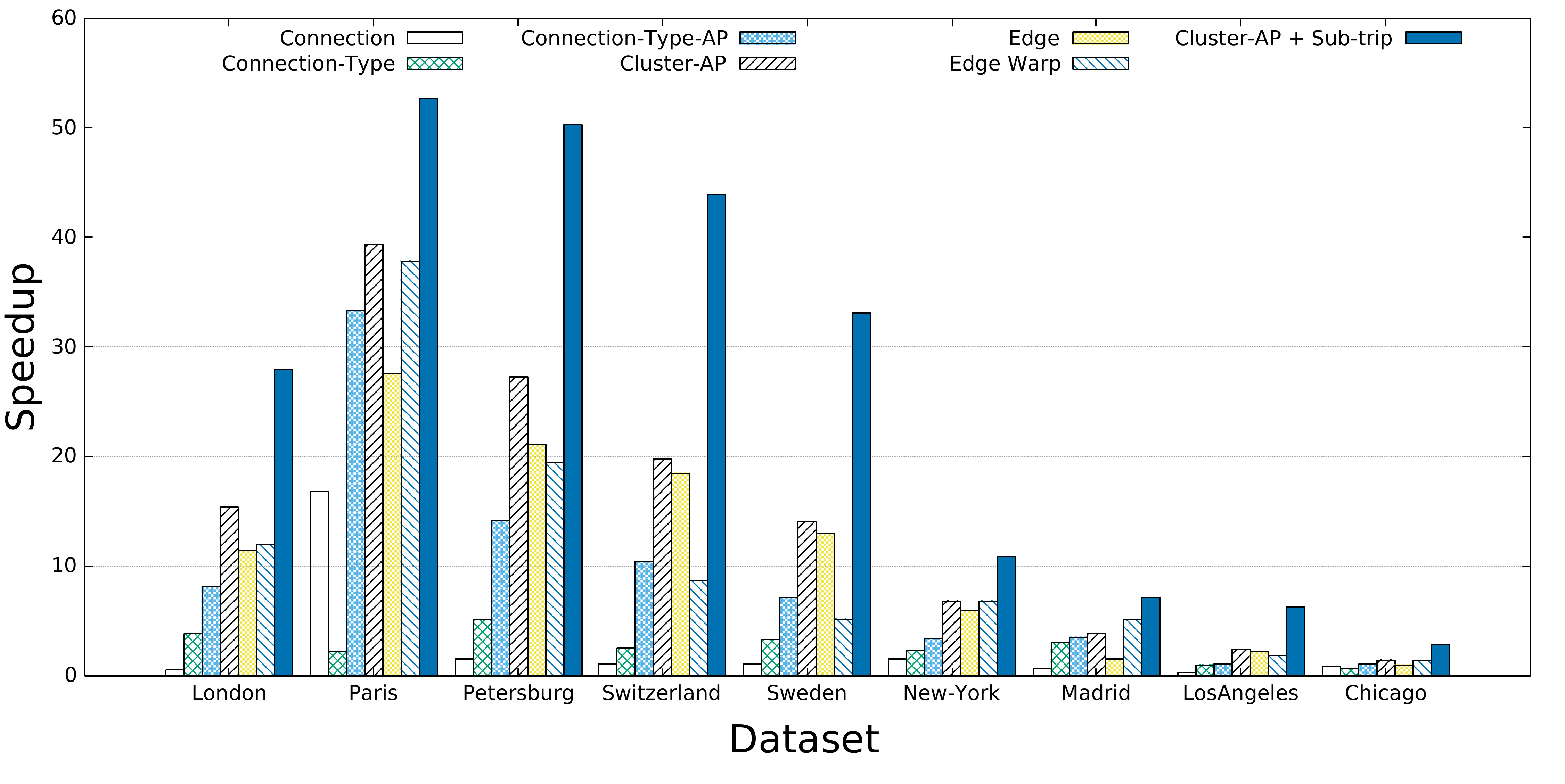}
    \vspace*{-5pt}
    \caption{Speedup achieved over connection-scan algorithm\label{fig:speedup}}
    \vspace*{-10pt}

\end{figure}

We run a thousand number of queries for each data-set on
the proposed algorithms and connection-scan algorithm, and
compute the average execution time.  Each query consists of a source vertex and a departure time. We choose a hundred source vertices randomly and for every source vertex, we again choose ten departure times uniformly at random for creating a thousand queries.
The average running times of the proposed algorithms are given in Table~\ref{executionTime},
and the speedups of our approaches with respect to the connection-scan algorithm are shown in Fig.~\ref{fig:speedup}.
In most of the data-sets, we can observe an incremental speedup among the algorithm versions namely Connection, Connection-type, Connection-type \textsc{ap} and Cluster-\textsc{ap} as these are developed with incremental improvements. We develop the Cluster-\textsc{ap} technique by combining Connection-type, clustering and arithmetic progression with the goal of pruning many irrelevant connections and decrease the running time. The Cluster-\textsc{ap} algorithm performs best among the proposed parallel algorithms.
As seen in the Fig.~\ref{fig:speedup}, the Cluster-\textsc{ap} algorithm with sub-trip gives the best performance across all the versions. In this version, we enhance the data using sub-trips and then run the Cluster-AP algorithm on the enhanced data. 

Also, the speedup varies between each data-set, for instance, the speedup of Cluster-\textsc{ap} algorithm on Paris data-set is $39.40\times$ whereas on Madrid data-set it is $3.77\times$. We claim that the speedups are not consistent on various data-sets due to their quality. 
Now, we shall look at a metric to measure the quality of a data-set, and observe the claim.

\begin{table*}
\caption{Execution Time in milliseconds. \label{executionTime}}
\centering
\begin{tabular}{|l|c|c|c|c|c|c|c|c|}
\hline
\textbf{Dataset} & \textbf{Serial} & \textbf{Connection} & \textbf{\begin{tabular}[c]{@{}c@{}}Connection\\ -type\end{tabular}} & \textbf{\begin{tabular}[c]{@{}c@{}}Connection\\ -type-AP\end{tabular}} & \textbf{Cluster-\textsc{ap}} & \textbf{Edge} & \textbf{Warps} & \textbf{\begin{tabular}[c]{@{}c@{}}Cluster-\textsc{ap} \\ + Sub-trips\end{tabular}} \\ \hline
London & 89.35 & 165.92 & 23.57 & 11.02 & 5.82 & 7.87 & 7.48 & 3.20 \\ \hline
Paris & 4.95 & 0.30 & 2.36 & 0.15 & 0.13 & 0.18 & 0.13 & 0.09 \\ \hline
Petersburg & 23.72 & 15.42 & 4.67 & 1.67 & 0.87 & 1.12 & 1.22 & 0.47 \\ \hline
Switzerland & 53.11 & 48.61 & 21.05 & 5.08 & 2.68 & 2.87 & 6.18 & 1.21 \\ \hline
Sweden & 40.57 & 36.74 & 12.52 & 5.70 & 2.90 & 3.13 & 7.85 & 1.22 \\ \hline
New-York & 2.48 & 1.70 & 1.08 & 0.73 & 0.36 & 0.42 & 0.36 & 0.23 \\ \hline
Madrid & 11.58 & 19.99 & 3.88 & 3.29 & 3.07 & 7.85 & 2.27 & 1.64 \\ \hline
Los Angeles & 11.62 & 36.24 & 11.86 & 11.32 & 4.86 & 5.33 & 6.39 & 1.86 \\ \hline
Chicago & 0.60 & 0.77 & 1.08 & 0.58 & 0.42 & 0.63 & 0.45 & 0.22 \\ \hline
\end{tabular}
\end{table*}

For a temporal graph $G$, let $\tilde{G}$ denote the edge-dependency graph of $G$, $|C|$ denotes the number of connections in $G$, $l(\tilde{G})$ denotes the maximum length of a longest path in $\tilde{G}$, $d(G)$ denote the temporal diameter of $G$. The edge-dependency graph based algorithm runs for $l(\tilde{G})$ iterations and the average number of connections processed per iteration is $|C|/l(\tilde{G})$, and thus the \textit{ parallel factor} of $G$ is defined as $\frac{|C|}{l(\tilde{G})}$ \cite{parallelAlgorithmForEAT}.
Our algorithm runs for atmost $d(G)$ iterations. 
We observe that $d(G) \leq l(\tilde{G})$ and thus the \emph{theoretical parallel factor} $p(G)$ is defined as $\frac{|C|}{d(G)}$ in this work.
Table~\ref{table_parallelism_factor} shows the theoretical parallel factor for each data-set. We provide an insight that for each data-set, the speedup of the proposed algorithms and $P$ are correlated. The data-sets London, Paris, Petersburg, Switzerland, and Sweden have a higher parallel factor and hence their speedups are higher than other data-sets. Also, for the data-sets Chicago, Los Angeles, Madrid and NewYork the parallel factor and the speedups are low.  

\begin{table}

\caption{Theoretical parallel factor for data-sets.\newline($K = 10^3$, $M=10^6$) \label{table_parallelism_factor}}
\centering
\begin{tabular}{|l|c|c|c|}
\hline
\textbf{Dataset $G$} & \textbf{$|C|$} & \textbf{$d(G)$} & \textbf{$p(G)$} \\ \hline
London & 14.06M & 283 & 49.68K \\ \hline
Paris & 1.06M & 30 & 35.33K \\ \hline
Petersburg & 4.43M & 190 & 23.32K \\ \hline
Switzerland & 9.26M & 418 & 22.15K \\ \hline
Sweden & 6.56M & 323 & 20.31K \\ \hline
New-York & 514.39K & 64 & 8.04K \\ \hline
Madrid & 1.99M & 232 & 8.58K \\ \hline
Los Angeles & 1.97M & 385 & 5.12K \\ \hline
Chicago & 98.15K & 47 & 2.09K \\ \hline
\end{tabular}

\end{table}




\subsection{Parameter Tuning}
The effect of cluster size, sub-trip size and virtual-warp size on the running time of the best algorithm Cluster-\textsc{ap} is analyzed in this section.

\begin{figure}
\centering
\includegraphics[width = 0.445\textwidth, keepaspectratio]{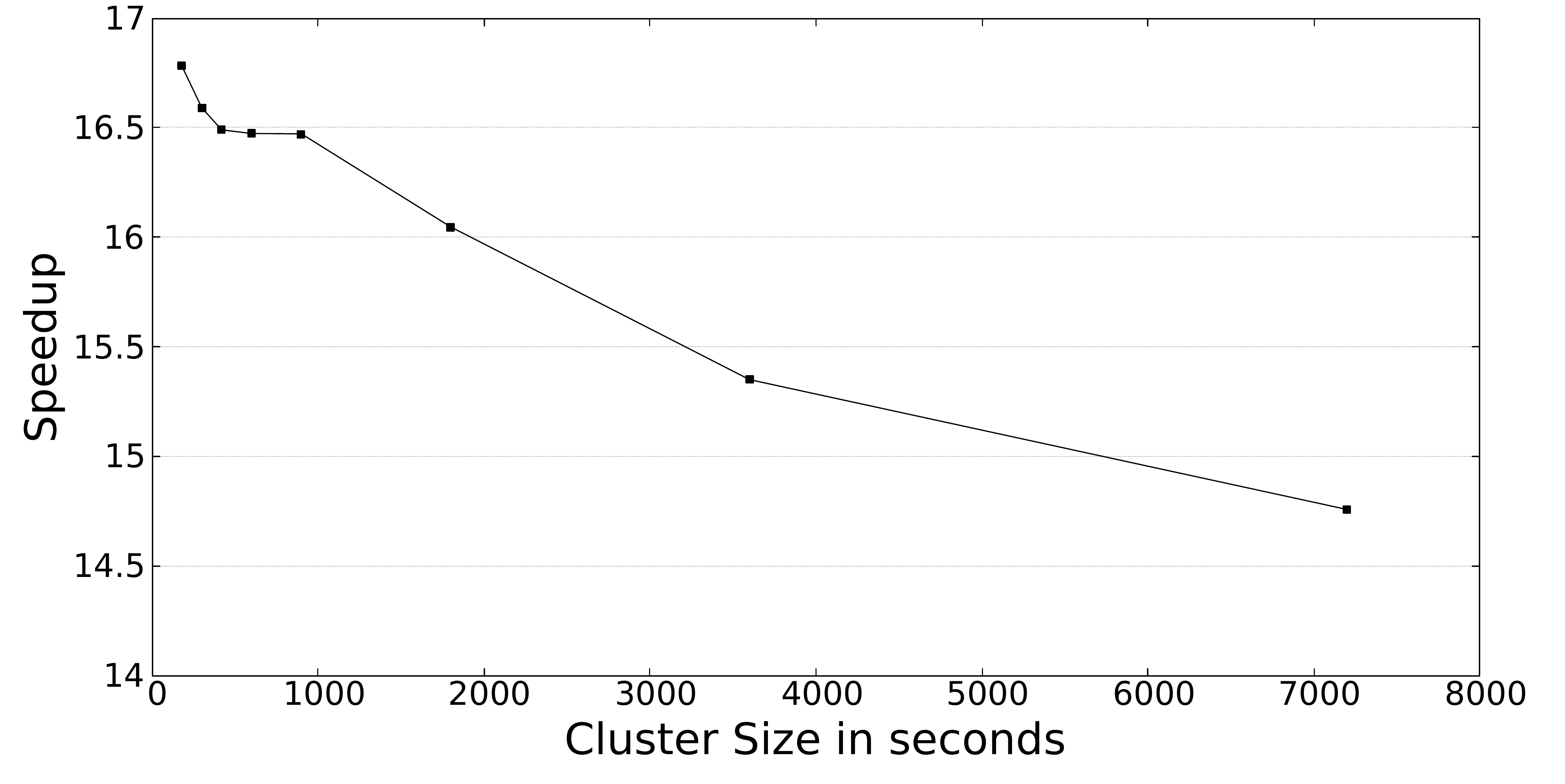} 

\caption{ Speedup of Cluster-\textsc{ap} over \textsc{csa} on London data-set for various cluster sizes.}\label{clusterAPAnalysis}
\vspace*{-10pt}
\end{figure}

\noindent
\textbf{Cluster Size Analysis:}
In the Cluster-\textsc{ap} algorithm, connections in a connection-type are partitioned into 24 clusters, such that the connections whose duration time lies between \texttt{i:00:00} and \texttt{i:59:59} are in the $i^{th}$ cluster. In other words, a connection $c$ belongs to $i^{th}$ cluster if and only if the hour information of the duration time of $c$ is $i$. The size of each cluster in a partition is referred as \emph{cluster size}, and the cluster size  in the above partitioning is one hour (60 minutes). 
We experiment the process of partitioning with various cluster sizes such as 30 minutes, 15 minutes, and 5 minutes, and compute the speedups. We perform this analysis on the London data-set as the number of connections and the parallelism factor is more when compared to the rest of the data-sets.
Fig.~\ref{clusterAPAnalysis} shows the effect of various cluster sizes on the speedup of the Cluster-\textsc{ap} algorithm. 


Note that the total number of connections per connection-type is constant. Therefore, as the cluster size decreases, the number of clusters increases and the number of connections per cluster decreases; also the search space for the \textsc{getConnectionFromAPs} function in the Cluster-\textsc{ap} algorithm decreases. Thus the speedup can be increased by reducing the cluster size of the partition and the evidence is shown in Fig.~\ref{clusterAPAnalysis}.

\noindent
\textbf{Sub-trips Size Analysis:}   The data-enhancement technique \emph{sub-trips} described in section \ref{subtrips} reduces the temporal diameter of the input graph and decreases the running time of proposed algorithms. A sub-trip of a trip is a non-overlapping sub-sequence of consequent connections. We divide the trip into sub-trips having equal number of connections and add artificial connections between the endpoints of the sub-trip. Adding these artificial connections creates shortcuts in the graph. The number of vertices skipped by these shortcuts is proportional to the length of sub-trips.
    Also, the length of sub-trips determines the total number of shortcuts. 
    As the amount of work is proportional to the number of connections, adding new shortcut connections increases the amount of work. Thus, we analyze the behavior of the data-enhancement technique by varying the lengths of sub-trips. 
    
    
    We experiment the following approaches to select a reasonable length for sub-trips.
        In the first approach, for each trip having trip-length $k$, we partition the trips into sub-trips having length $\sqrt{k}$. Thus for every trip, the number of sub-trips is made as equal to the length of sub-trips. We now describe the drawbacks of the first approach. The trips having shorter length has no useful effect of shortcuts as they skip a smaller number of vertices. Such shortcuts are more of an overhead than enhancement. Also, the trips with larger lengths have large sub-trips. The difference in the lengths of sub-trips causes unfairness. For instance, consider two trips $t_1$ and $t_2$ whose lengths are  9 and 100 respectively. We add shortcuts to $t_1$ and $t_2$ where the sub-trips lengths are 3 and 10, respectively. It can be seen that
        the length of the sub-trips in $t_2$ is greater than the trip-length of $t_1$.
        In reality, $t_2$ requires more sub-trips compared to $t_1$ and the following scheme overcomes this drawback.
        We first compute the average trip length $r$ over all the trips in the public transport data, and
        divide each trip into sub-trips of lengths $\sqrt{r}$.

        We run the Cluster-\textsc{ap} algorithm on London data-set before and after data-enhancement and find speedup when compared to the connection-scan algorithm. Before data-enhancement, the speedup is $15.35\times$, whereas the speedup is $27.88 \times$ and $28.10\times$ on enhanced data using the first approach and second approach, respectively. 
        As shown in Table~\ref{table_parallelism_factor_subtrip}, the number of connections in the enhanced data is increased when compared to the original data. However, the temporal diameter is decreased and the parallel factor is increased in the enhanced data.
        Thus the data-enhancement technique helps to increase the data quality and decreases the running time.
        


\begin{table}

\caption{Theoretical parallel factor for enhanced data-sets.\newline($K = 10^3$, $M=10^6$) \label{table_parallelism_factor_subtrip}}
\centering
\begin{tabular}{|l|c|c|c|}
\hline
\begin{tabular}[c]{@{}l@{}}\textbf{Enhanced}\\ \textbf{Dataset $G$}\end{tabular} & $|C|$ & $d(G)$ & $p(G)$ \\ \hline
London & 15.92M & 127 & 125.38K \\ \hline
Paris & 1.50M & 16 & 93.67K \\ \hline
Petersburg & 5.40M & 78 & 69.29K \\ \hline
Switzerland & 11.35M & 146 & 77.75K \\ \hline
Sweden & 7.76M & 96 & 80.81K \\ \hline
New-York & 591.94K & 30 & 19.73K \\ \hline
Madrid & 2.36M & 71 & 33.28K \\ \hline
Los Angeles & 2.21M & 109 & 20.31K \\ \hline
Chicago & 120.65K & 19 & 6.35K \\ \hline
\end{tabular}

\end{table}

\noindent
\textbf{Warps Size Analysis:}
\begin{figure}
    \centering
    \includegraphics[width=0.3\textwidth]{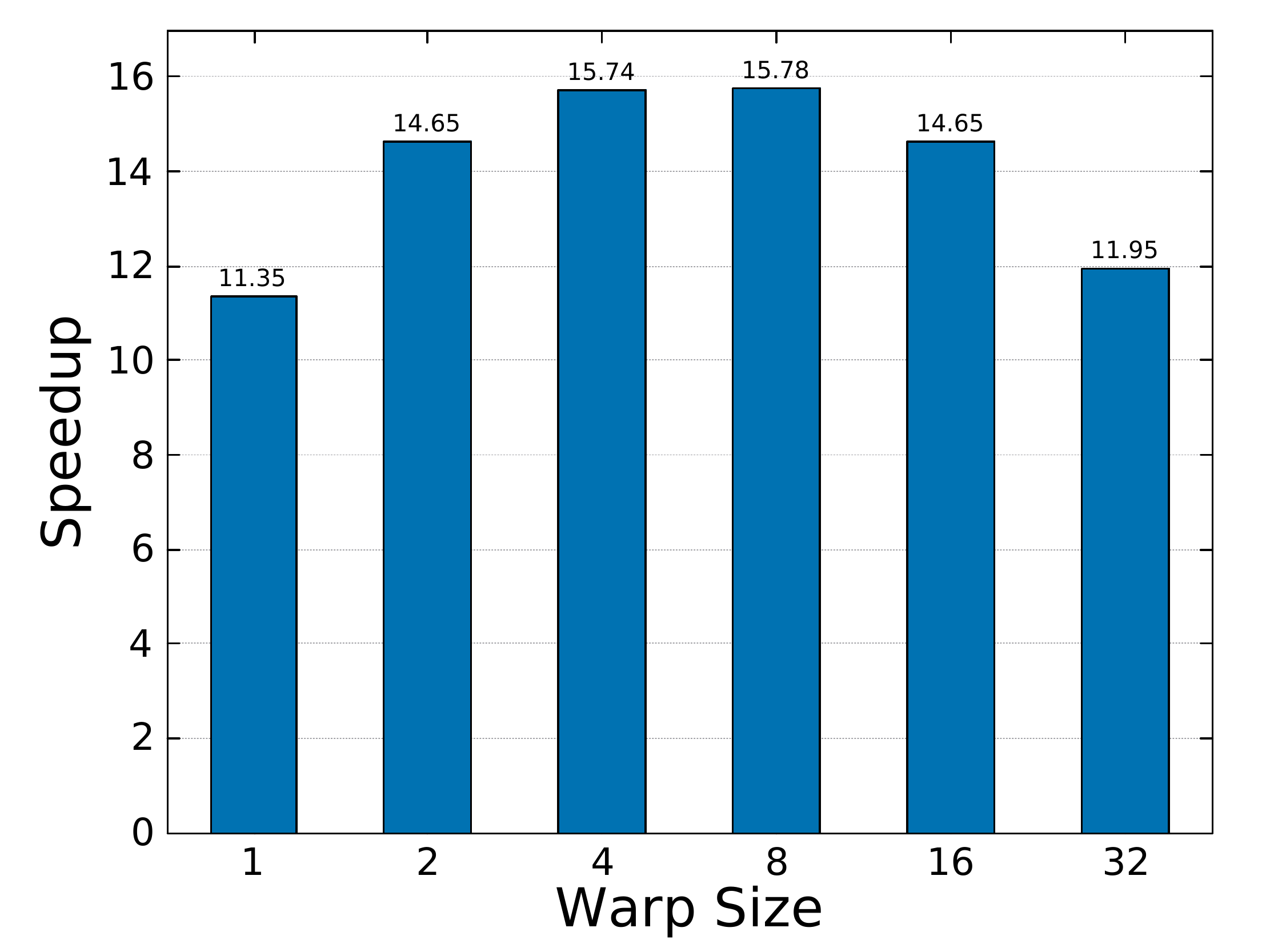}
    \caption{ Speedup of Edge-warp Algorithm over \textsc{csa} on London data-set for various warp sizes.\label{warpExperiment}}
    \vspace*{-10pt}
\end{figure}
In section \ref{warps}, we proposed a warp-centric Edge-version algorithm. The key idea is to allocate warps with edges. This mapping ensures that threads working in the same warp do not diverge. Also, warps help to reduce the number of scattered memory accesses. The only problem in allocating warp to an edge is under-utilization of threads. Some edges may have very few connection-types which causes the threads in the warp under-utilized. To resolve this, the threads in warp can be partitioned into sub-warps of equal sizes, which are referred to as virtual warps \cite{virtualWarp}. These sub-warps are then mapped to multiple edges. Hence, a warp works on multiple edges, where each sub-warp processes an edge. The size of a sub-warp is a parameter which we rehearse with various values. 
The Fig.~\ref{warpExperiment} shows the speedup of warp-centric Edge-version algorithm compared to \textsc{csa} for various sizes of virtual warps.
 If the sub-warp size is made as one it is equivalent to the edge-version as every thread is mapped to an edge. 

    Adding sub-warps introduces some thread-divergence. However, the number of scattered memory accesses is limited to the number of sub-warps and not the total number of threads in a warp.  Thus, the running time of our algorithm is reduced by managing the thread divergence, scattered memory accesses, and the proper utilization of threads in a warp effectively.
    

\subsection{Comparison with ESDG algorithm}
In this section, we compare the performance of Cluster-\textsc{ap} algorithm with the \textsc{esdg} algorithm implemented on \textsc{gpu}. Further, we propose an optimization technique to reduce the memory copy between \textsc{cpu} and \textsc{gpu} in the proposed algorithms. This technique helps to achieves 1.63$\times$ to 12.48$\times$ times speedup over the \textsc{esdg} algorithm.

As per \textsc{esdg} algorithm, the connections are arranged in increasing order based on their level number, and hence the number of connections in each level are known in the preprocessing time. 
In the \textsc{gpu} implementation of \textsc{esdg} algorithm, we launch $l_i$ threads in $i^{th}$ iteration to relax $l_i$ connections in level-$i$, and the algorithm terminates when the connections are relaxed at all levels. The Cluster-\textsc{ap} gives upto 5.21$\times$ speedup over the \textsc{esdg gpu} implementation. Referring to the \textsc{esdg} parallel algorithm, we realize that there is a memory copy operation between \textsc{cpu} and \textsc{gpu}, which adds memory copy penalty in  all iterations in the proposed algorithms. The general structure of \textsc{gpu}-implementation of our algorithm is as follows:

\begin{algorithmic}[1]
\While{flag} \Comment{\textsc{cpu}}
\State{flag = $false$} \Comment{\textsc{cpu}}
\State{\textsc{copyCputoGpu}(flag,dFlag)} \Comment{\textsc{cpu} and \textsc{gpu}}
\For{}\textbf{ in parallel} \Comment{\textsc{gpu}}
    \State{\textcolor{darkgray}{/* do work */}} \Comment{\textsc{gpu}}
    \State dFlag = $true$ \textbf{or} $false$ \Comment{\textsc{gpu}}
\EndFor
\State{\textsc{copyGputoCpu}(dFlag,flag)} \Comment{\textsc{cpu} and \textsc{gpu}}
\EndWhile
\end{algorithmic}

Lines 3 and 6 in the above pseudo-code cause a memory copy penalty in every iteration. To avoid this penalty, we perform memory copy mentioned in Lines 3 and 6 in $i^{th}$ iteration, where $i\in\{0, \sqrt{d}, 2\sqrt{d},\ldots,d\}$ and $d$ is the temporal diameter of the graph. We provide the execution times of the algorithms in table \ref{executionTimeESDG}. This implementation gives a speedup in the range of 1.63$\times$ to 12.48$\times$ over \textsc{esdg} across the datasets. 
The Cluster-\textsc{ap} algorithm processes on average only 3.35\% of the connections whereas \textsc{esdg} processes all the connections in the graph.
Also, the \textsc{esdg} performance is query independent, i.e., regardless of a query, the algorithm performs the same amount of work, whereas our algorithms benefit from pruning and early termination techniques.



\begin{table}
\caption{Execution Time in milliseconds\label{executionTimeESDG}}
\centering
\begin{tabular}{|l|c|c|c|}
\hline
\textbf{Dataset} & \textbf{ESDG} & \textbf{Cluster-\textsc{ap}} & \textbf{\begin{tabular}[c]{@{}c@{}}Cluster-\textsc{ap} + (Reduced \\  Memory Transfers)\end{tabular}} \\ \hline
London & 6.23 & 5.82 & 3.80 \\ \hline
Paris & 0.27 & 0.13 & 0.08 \\ \hline
Petersburg & 5.06 & 0.87 & 0.40 \\ \hline
Switzerland & 5.66 & 2.68 & 1.59 \\ \hline
Sweden & 6.65 & 2.90 & 1.88 \\ \hline
New-York & 2.13 & 0.36 & 0.21 \\ \hline
Madrid & 4.73 & 3.07 & 2.27 \\ \hline
Los Angeles & 5.70 & 4.86 & 2.18 \\ \hline
Chicago & 2.01 & 0.42 & 0.27 \\ \hline
\end{tabular}
\end{table}

    
    
\section{Related Work}
    


Xuan et.al, have initiated the foremost journey problem \cite{foremostJourney2003} and is referred to as earliest arrival path in the recent work \cite{pathProblemsInTemporalGraphs}.
Every prefix path of a shortest path is a shortest path, whereas every prefix path of an earliest arrival path need not be an earliest arrival path. However, it was shown that if there exists an earliest arrival path $P$ from a vertex $s$ to a vertex $t$, then every prefix path of $P$ is an earliest arrival path. This property is used to design a variant of Dijkstra's algorithm using priority-queue for solving single-source \textsc{eat} \cite{foremostJourney2003}.
The variant of Dijkstra's algorithm requires the data to be in a graph format, whereas the connection-scan algorithm to solve single source \textsc{eat} works on the array of connections without any graph.
In the single source \textsc{eat} problem, the earliest arrival times are computed from the single source to the rest of the vertices, whereas the earliest arrival time from the given source to the given destination is computed in the goal-directed \textsc{eat} problem.
The classical algorithms to solve goal-oriented public transport based problems such as \textsc{eat} problem, profile-search problem, and the multi-criteria problem are \textsc{raptor}, Transfer patterns, and Trip-based algorithm \cite{raptor,transferPatterns,tripBased}. These algorithms are not specifically designed for single-source \textsc{eat}. 
The \textsc{raptor} algorithm does not require any preprocessing whereas the other two algorithms require heavy preprocessing.
Transfer pattern and trip based egardless  of  a  query, algorithms maintain directed acyclic graphs and prefix trees for each vertex, respectively. 
A multi-core parallel algorithm was designed to compute all non-dominated paths (profile-search problem), from a single source to all the vertices \cite{parallelProfileSearch}. In this work, 
the parallelism is exploited at a source vertex, i.e.,
all the out-going connections from the source vertex are divided into $p$ groups, where $p$ is the available number of processors, and $i^{th}$ processor runs a serail algorithm from the source vertex by considering the connections from $i^{th}$ group.
The edge-scan-dependency graph algorithm was proposed to solve single source \textsc{eat} using multi-core processors, in which parallelism is applied to all the connections available at the same level. In our Cluster-AP algorithm, the parallelism is exploited to process all the connection-types at a time.
Our algorithm prunes many connections from each connection-type,
whereas the edge-dependency graph algorithm processes all the connections.

\section{Conclusion}
We have designed a hierarchical data-structure and efficient searching techniques to implement the topological driven parallel algorithm on the graphics processing unit, for solving the single source earliest arrival time problem on public transport networks.
In our data-structure, all the connections associated with each edge are divided into groups in such a way that the connections that are having the same duration are kept in the same group. 
Further, the connections within each group are divided into clusters in such a way that the connections whose departure time lies in the $i^{th}$ hour of a day are arranged in $i^{th}$  cluster. Later, the departure times of all connections in each cluster are compactly represented as a sequence of arithmetic progressions. During the traversal of an edge from $u$ to $v$, our algorithm makes use of this data-structure to retrieve a connection from $u$ to $v$ through which $v$ can be reached earliest, without processing many connections associated with the edge. 
We have examined the running times of the best implementation of our Cluster-\textsc{ap} algorithm, against the serial connection-scan algorithm and the \textsc{gpu}-implementation of the edge-scan-dependency-graph algorithm on various data-sets, and the average speedups range from $2.29\times$ to $59.09\times$ and $1.63\times$ to $12.48\times$, respectively.
The primary reason for this accomplishment is that, our algorithm processes around $471$ thousand connections out of $14$ million connections on London-data set.
Our pruning techniques and data-structure would benefit to solve many other problems in public-transport networks and general temporal graphs.

\bibliographystyle{ieeetr}
\bibliography{eat1}

\begin{thebibliography}{10}

\bibitem{IntriguinglyPaper}
J.~Dibbelt, T.~Pajor, B.~Strasser, and D.~Wagner, ``Intriguingly simple and
  fast transit routing,'' in {\em Proc. of the SEA}, pp.~43--54, June 2013.

\bibitem{pathProblemsInTemporalGraphs}
H.~Wu, J.~Cheng, S.~Huang, Y.~Ke, Y.~Lu, and Y.~Xu, ``Path problems in temporal
  graphs,'' {\em VLDB Endowment}, vol.~7, pp.~721--732, May 2014.

\bibitem{parallelAlgorithmForEAT}
P.~{Ni}, M.~{Hanai}, W.~J. {Tan}, C.~{Wang}, and W.~{Cai}, ``Parallel algorithm
  for single-source earliest-arrival problem in temporal graphs,'' in {\em
  Proc. of the ICPP}, pp.~493--502, Aug 2017.

\bibitem{foremostJourney2003}
B.~Bui{-}Xuan, A.~Ferreira, and A.~Jarry, ``Computing shortest, fastest, and
  foremost journeys in dynamic networks,'' {\em Int. J. Found. Comput. Sci.},
  vol.~14, no.~2, pp.~267--285, 2003.

\bibitem{gpuBook}
D.~B. Kirk and W.~H. Wen-Mei, {\em Programming massively parallel processors: a
  hands-on approach}.
\newblock Morgan kaufmann, 2016.

\bibitem{biconnectedComponents}
M.~Wadwekar and K.~Kothapalli, ``A fast {GPU} algorithm for biconnected
  components,'' in {\em Proc. of the {IC3}}, pp.~1--6, August 2017.

\bibitem{stronglyConnectedCompoenents}
S.~Devshatwar, M.~Amilkanthwar, and R.~Nasre, ``{GPU} centric extensions for
  parallel strongly connected components computation,'' in {\em Proc. of the
  9th Annual Workshop on GPGPU, PPoPP}, pp.~2--11, March 2016.

\bibitem{frequencyBasedPaper}
H.~Bast and S.~Storandt, ``Frequency-based search for public transit,'' in {\em
  Proc. of the Int. Conf. 22Nd ACM SIGSPATIAL}, pp.~13--22, 2014.

\bibitem{virtualWarp}
S.~Hong, S.~K. Kim, T.~Oguntebi, and K.~Olukotun, ``Accelerating {CUDA} graph
  algorithms at maximum warp,'' in {\em Proc. of the 16th {ACM} {SIGPLAN},
  {PPOPP}}, pp.~267--276, February 2011.

\bibitem{raptor}
D.~Delling, T.~Pajor, and R.~F. Werneck, ``Round-based public transit
  routing,'' {\em Transportation Science}, vol.~49, no.~3, pp.~591--604, 2015.

\bibitem{transferPatterns}
H.~Bast, E.~Carlsson, A.~Eigenwillig, R.~Geisberger, C.~Harrelson, V.~Raychev,
  and F.~Viger, ``Fast routing in very large public transportation networks
  using transfer patterns,'' in {\em Proc. of the {ESA} -Algorithms,Part {I}},
  pp.~290--301, September 2010.

\bibitem{tripBased}
S.~Witt, ``Trip-based public transit routing,'' in {\em Proc. of the {ESA}
  -Algorithms,}, pp.~1025--1036, September 2015.

\bibitem{parallelProfileSearch}
D.~Delling, B.~Katz, and T.~Pajor, ``Parallel computation of best connections
  in public transportation networks,'' {\em Journal of Experimental
  Algorithmics}, vol.~17, pp.~1--12, 2012.

\end{thebibliography}

\end{document}